# Hollow Gaussian Schell-model beam and its propagation

Li-Gang Wang [*]

*Department of Physics, Zhejiang University, Hangzhou, 310027, China*

**Abstract**

In this paper, we present a new model, hollow Gaussian-Schell model beams (HGSMBs), to describe the practical dark hollow beams. An analytical propagation formula for HGSMBs passing through a paraxial first-order optical system is derived based on the theory of coherence. Based on the derived formula, an application example showing the influence of spatial coherence on the propagation of beams is illustrated. It is found that the beam propagating properties of HGSMBs will be greatly affected by their spatial coherence. Our model provides a very convenient way for analyzing the propagation properties of partially coherent dark hollow beams. © 2007 Elsevier Science B. V. All rights reserved.

## 1. Introduction

Recently, dark hollow beams, which are also called all-light guides (ALGs), have been paid a lot of attention [1-6] due to their potential applications in atom optics. These ALGs are very useful tools for atom lithography [2], atom interferometry [7], and atomic spectroscopy, and also for transporting and manipulating Bose-Einstein condensates of atoms [3, 8-9] and micro-sized particles [10]. There are various of methods for generating a dark hollow beam, such as the mode selection method by designing a particular laser cavity [11-12], the computer-generated hologram method with an interference pattern recorded on a film [13], the geometrical optical method by combining of three axicons and a simple lens [8], and the methods by using the vortex gratings [14], hollow-core fiber [15], refractive conical lenses [16], and self-phase modulation [17], and so on.

---

[*] E-mail address: sxwlg@yahoo.com.cn

Theoretically, several models are introduced to describe dark hollow beams. One of them is a $TEM_{01}^*$ mode beam, this is the simplest model. There are also some other models to describe dark hollow beams, such as the high-order Laguerre-Gaussian beams [4, 18-19], high-order Bessel beams [20], and controllable dark-hollow beams [5, 21-22]. Recently, Zhang et al. presented a new model named hollow Gaussian beams (HGBs) to describe the dark hollow beam, and they also experimentally verified that such HGBs can be generated by designing the laser cavity with diffraction optical elements [12]. It has been proven that HGBs can be expressed as a superposition of a series of Lagurere-Gaussian modes [5]. However, all previous investigations have been limited within the fully coherent case, and most practical beams are partially coherent because any completely coherent beam does not exist in the strict sense. Furthermore, in practice, multimode oscillation always results in partial coherent [23-24]. As yet, the propagation properties of partially coherent dark hollow beams are little studied, then the problem arise here is that how the spatial coherent of light beam affects on the propagation properties of the dark hollow beams. In this communication, we propose a new model named hollow Gaussian-Schell model beams (HGSMBs) to describe the practical dark hollow beams. We derive a general propagating formula of the HGSMB through a paraxial *ABCD* optical system. It is found that the propagating properties of HGSMBs will be greatly affected by their spatial coherence. Our numerical example indicates that the range of dark-hollow region will be greatly affected by the coherence of the beam.

**2. Definition of HGSMBs**

For a coherent HGB, in rectangular coordinates its electric field at $z=0$ could be defined by

$$E(x,y,0) = G_0 \left(\frac{x^2+y^2}{\sigma_I^2}\right)^n \exp\left[-\frac{x^2+y^2}{4\sigma_I^2}\right], \tag{1}$$

where the integer $n = 0,1,2,...$ is the order of the HGB, and $G_0$ is a constant. For $n=0$, Eq. (1) reduces to a fundamental Gaussian beam with the beam half-width $\sigma_I$. For $n=1,2,3,\cdots$, there is a dark region at the center area of the transverse profile and this dark region could be controlled by changing the integer $n$, the radius of the bright ring of the HGB increases as the integer $n$ increases, so does the area of the dark region of the HGBs.

Based on the theory of coherence, the cross-spectral density for a partially coherent beam generated by a Schell-model source at the initial plane $z=0$ can be expressed in the following well-known form [25]:

$$W(x_1, y_1, x_2, y_2; 0) = [I(x_1,y_1,0)]^{1/2}[I(x_2,y_2,0)]^{1/2}\mu(x_1-x_2, y_1-y_2, 0), \tag{2}$$

where $I(x_i, y_i, 0)$ ($i=1,2$) are the intensity distribution of the source, and $\mu(x_1-x_2, y_1-y_2, 0)$ is the spectral degree of coherence given by

$$\mu(x_1-x_2, y_1-y_2, 0) = e^{-[(x_1-x_2)^2+(y_1-y_2)^2]/(2\sigma_g^2)}. \tag{3}$$

Here $\sigma_g$ is called the transverse coherence width of the source, and it denotes the spatial coherence. From Eq. (1), we can obtain the intensity distribution for the coherent HGB given by

$$I(x,y,0) = E(x,y,0)E^*(x,y,0) = G_0^2\left(\frac{x^2+y^2}{\sigma_I^2}\right)^{2n} \exp\left[-\frac{x^2+y^2}{2\sigma_I^2}\right]. \tag{4}$$

From Eq. (2), it is obvious that the intensity distribution (at the initial plane) for a partially coherent Schell-model beam [25] is the same for the beams with different spatial coherence (i. e., different values of $\sigma_g$). Therefore, from Eq. (2), (3) and (4), the cross-spectral density of partially coherent HGBs can be expressed as follows:

$$W(x_1, y_1, x_2, y_2; 0) = G_0^2[(x_1^2+y_1^2)/\sigma_I^2]^n[(x_2^2+y_2^2)/\sigma_I^2]^n e^{-(x_1^2+y_1^2)/(4\sigma_I^2)}e^{-(x_2^2+y_2^2)/(4\sigma_I^2)} \\ \times e^{-[(x_1-x_2)^2+(y_1-y_2)^2]/(2\sigma_g^2)}. \tag{5}$$

Equation (5) describes the cross-spectral density of a partially coherent HGB at the initial plane. From Eq. (5), we easily find that the intensity profiles of partially coherent HGBs at $z=0$ are the same when $\sigma_g$ changes from zero (fully incoherent) to infinity (fully coherent). Similar to the Gaussian Schell-model beams [25], the beams described by Eq. (5) may be called as *hollow Gaussian Schell-model beams* (HGSMBs). Therefore, Eq. (5) may give an explicit expression for describing the practical dark hollow beams with a controllable dark region and the adjustable spatial coherence.

3 **Propagating formula of HGSMBs**

Now let us consider the propagation of a HGSMB passing through a paraxial *ABCD* optical system. It is well known that the propagation of a coherent beam passing through a paraxial *ABCD* optical can be treated by the Collins formula [26-27] as follows:

$$E(u,v,z) = -\frac{ik}{2\pi B} e^{ikL_0} \iint E(x,y,0) \exp\left\{\frac{ik}{2B}\left[A(x^2+y^2) - 2(xu+yv) + D(u^2+v^2)\right]\right\} dxdy, \quad (6)$$

where $k = 2\pi/\lambda$ is the wave number, $\lambda$ is the wavelength, and $A$, $B$ and $D$ denote the transfer matrix elements of the paraxial optical system.

For the partially coherent fields, we can assume that the fields at the two arbitrary points $(x_1, y_1)$ and $(x_2, y_2)$ in the incident plane $z=0$ are $E(x_1, y_1)$ and $E(x_2, y_2)$, respectively, and the fields at two arbitrary points $(u_1, v_1)$ and $(u_2, v_2)$ in the output plane $z$ are $E(u_1, v_1)$ and $E(u_2, v_2)$, respectively. Then the cross-spectral density in the incident and output planes could be expressed by [25] $W(x_1, y_1, x_2, y_2; 0) = \langle E(x_1, y_1, 0) E^*(x_2, y_2, 0) \rangle$ and $W(u_1, v_1, u_2, v_2; z) = \langle E(u_1, v_1, z) E^*(u_2, v_2, z) \rangle$, where $\langle \ \rangle$ denotes an ensemble average. Using Eq. (6), it is easy to find that the propagation formula for the cross-spectral density of a partially coherent beam through a paraxial *ABCD* optical system as follows:

$$W(u_1, v_2, u_2, v_2; z) = \langle E(u_1, v_1, z) E^*(u_2, v_2, z) \rangle$$
$$= \frac{k^2}{4\pi^2 B_1 B_2} \iiiint W(x_1, y_1, x_2, y_2; 0) \exp\left\{ \frac{ik}{2B_1} \left[ A_1(x_1^2 + y_1^2) - 2(x_1 u_1 + y_1 v_1) + D_1(u_1^2 + v_1^2) \right] \right\} \quad (7)$$
$$\times \exp\left\{ -\frac{ik}{2B_2} \left[ A_2(x_2^2 + y_2^2) - 2(x_2 u_2 + y_2 v_2) + D_2(u_2^2 + v_2^2) \right] \right\} dx_1 dy_1 dx_2 dy_2.$$

Here we have assumed the paraxial optical system to be stationary stable in the above equation.

Using the binomial theorem, the factors $[(x_i^2 + y_i^2)/\sigma_I^2]^n$ can be expanded into the summation form [28]: $\sum_{t=0}^{n} \binom{n}{t} x_i^{2(n-t)} y_i^{2t}$, where $i = 1, 2$, and $\binom{n}{t}$ denotes a binomial coefficient. The cross-terms $\exp(\frac{x_1 x_2}{\sigma_g^2})$ and $\exp(\frac{y_1 y_2}{\sigma_g^2})$ in the function of the spectral degree of coherence [see Eq. (3)] can be expanded [28] into $\sum_{m=0}^{\infty} \frac{1}{m!} (\frac{x_1 x_2}{\sigma_g^2})^m$ and $\sum_{f=0}^{\infty} \frac{1}{f!} (\frac{y_1 y_2}{\sigma_g^2})^f$, respectively. Using these mathematical relations, we readily find that the cross-spectral density of the HGSMB in the initial plane can be expressed in the form:

$$W(x_1, y_1, x_2, y_2; 0) = \frac{G_0^2}{\sigma_I^{4n}} \sum_{t=0}^{n} \sum_{j=0}^{n} \sum_{m=0}^{\infty} \sum_{f=0}^{\infty} \binom{n}{t}\binom{n}{j} \frac{1}{m! f!} \frac{1}{\sigma_g^{2m+2f}} x_1^{2(n-t)+m} y_1^{2t+f} x_2^{2(n-j)+m} y_2^{2j+f} \quad (8)$$
$$\times \exp\left[ -a(x_1^2 + y_1^2 + x_2^2 + y_2^2) \right],$$

where $a = 1/(4\sigma_I^2) + 1/(2\sigma_g^2)$. On substituting Eq. (8) into Eq. (7), and applying the following integral transformations [29]:

$$\int_{-\infty}^{\infty} (ix)^\nu e^{-\alpha^2 x^2} e^{\mp ixy} dx = \pi^{1/2} 2^{-\nu/2} \alpha^{-\nu-1} e^{-y^2 \alpha^{-2}/8} D_\nu\left( \pm 2^{-1/2} \alpha^{-1} y \right), \quad (9a)$$

$$H_n(x) = 2^{n/2} e^{x^2/2} D_n(\sqrt{2} x), \quad (9b)$$

$$H_n(-x) = (-1)^n H_n(x), \quad (9c)$$

after the tedious calculation, we can obtain the expression for the cross-spectral density of the output HGSMB:

$$W(u_1,v_1,u_2,v_2;z) = \frac{G_0^2 k^2}{4B_1 B_2 \sigma_1^{4n}} \exp\left(-\frac{k^2 u_1^2}{4P_1 B_1^2} - \frac{k^2 v_1^2}{4P_1 B_1^2} - \frac{k^2 u_2^2}{4P_2 B_2^2} - \frac{k^2 v_2^2}{4P_2 B_2^2}\right) \exp\left[-\frac{ikD_1}{2B_1}(u_1^2+v_1^2) + \frac{ikD_2}{2B_2}(u_2^2+v_2^2)\right]$$

$$\times \sum_{t=0}^{n}\sum_{j=0}^{n}\sum_{m=0}^{\infty}\sum_{f=0}^{\infty} \binom{n}{t}\binom{n}{j}\frac{1}{m!}\frac{1}{f!}\frac{1}{\sigma_g^{2m+2f}} 4^{-(2n+m+f)}(P_1 P_2)^{-(2n+m+f+2)/2}$$

$$\times H_{2(n-t)+m}\left(\frac{ku_1}{2P_1^{1/2}B_1}\right) H_{2(n-j)+m}\left(\frac{ku_2}{2P_2^{1/2}B_2}\right) H_{2t+f}\left(\frac{kv_1}{2P_1^{1/2}B_1}\right) H_{2j+f}\left(\frac{kv_2}{2P_2^{1/2}B_2}\right),$$

(10)

where $P_1 = a - ikA_1/(2B_1)$, $P_2 = a + ikA_2/(2B_2)$, $D_n(x)$ is the parabolic cylinder function, and $H_n(x)$ is the Hermite polynomials. With the help of the extension form of Mehler's formula [30]:

$$\sum_{q=0}^{\infty} H_{q+r}(x) H_{q+s}(y) \frac{t^q}{q!} = (1-4t^2)^{-(r+s+1)/2} \exp\left[\frac{4xyt - 4(x^2+y^2)t^2}{1-4t^2}\right]$$

$$\times \sum_{k=0}^{\min(r,s)} 2^{2k} k! \binom{r}{k}\binom{s}{k} t^k H_{r-k}\left[\frac{x-2yt}{(1-4t^2)^{1/2}}\right] H_{s-k}\left[\frac{y-2xt}{(1-4t^2)^{1/2}}\right],$$

(11)

we can get the final expression of the cross-spectral density of the output HGSMB:

$$W(u_1,v_1;u_2,v_2,z) = \frac{G_0^2 k^2}{4^{2n+1} B_1 B_2 \sigma_1^{4n}(P_1 P_2)^{n+1}} \left(1 - \frac{1}{4\sigma_g^4 P_1 P_2}\right)^{-(2n+1)} \exp\left[-\frac{ikD_1}{2B_1}(u_1^2+v_1^2) + \frac{ikD_2}{2B_2}(u_2^2+v_2^2)\right]$$

$$\times \exp\left(-\frac{k^2 u_1^2}{4P_1 B_1^2} - \frac{k^2 v_1^2}{4P_1 B_1^2} - \frac{k^2 u_2^2}{4P_2 B_2^2} - \frac{k^2 v_2^2}{4P_2 B_2^2}\right)$$

$$\times \exp\left[\frac{\frac{\sigma_g^2 k^2 u_1 u_2}{B_1 B_2} - \left(\frac{k^2 u_1^2}{4P_1 B_1^2} + \frac{k^2 u_2^2}{4P_2 B_2^2}\right)}{\eta^2-1}\right] \exp\left[\frac{\frac{\sigma_g^2 k^2 v_1 v_2}{B_1 B_2} - \left(\frac{k^2 v_1^2}{4P_1 B_1^2} + \frac{k^2 v_2^2}{4P_2 B_2^2}\right)}{\eta^2-1}\right]$$

$$\times \sum_{t=0}^{n}\sum_{j=0}^{n}\sum_{\alpha=0}^{\min[2(n-t),2(n-j)]}\sum_{\beta=0}^{\min[2t,2j]} \binom{n}{t}\binom{n}{j}\binom{2(n-t)}{\alpha}\binom{2(n-j)}{\alpha}\binom{2t}{\beta}\binom{2j}{\beta}\alpha!\beta!2^{2\alpha+2\beta}\left(\frac{1}{2\eta}\right)^{\alpha+\beta}$$

$$\times H_{2(n-t)-\alpha}\left(\frac{\frac{\eta k u_1}{2P_1^{1/2}B_1} - \frac{k u_2}{2P_2^{1/2}B_2}}{(\eta^2-1)^{1/2}}\right) H_{2(n-j)-\alpha}\left(\frac{\frac{\eta k u_2}{2P_2^{1/2}B_2} - \frac{k u_1}{2P_1^{1/2}B_1}}{(\eta^2-1)^{1/2}}\right)$$

$$\times H_{2t-\beta}\left(\frac{\frac{\eta k v_1}{2P_1^{1/2}B_1} - \frac{k v_2}{2P_2^{1/2}B_2}}{(\eta^2-1)^{1/2}}\right) H_{2j-\beta}\left(\frac{\frac{\eta k v_2}{2P_2^{1/2}B_2} - \frac{k v_1}{2P_1^{1/2}B_1}}{(\eta^2-1)^{1/2}}\right).$$

(12)

where $\eta = 2\sigma_g^2(P_1P_2)^{1/2}$. Equation (12) is the general propagation formula for HGSMBs passing through the paraxial *ABCD* optical system. It provides a convenient and powerful tool for treating the propagation and transformation of HGSMBs and is very helpful for describing the partially coherent dark hollow beams when one considers the effect of spatial coherence on the beam propagation. Obviously it is easily to verify that when $n = 0$, Eq. (12) could be simplified into the expression for Gaussian Schell-Model beams [25]. It should be pointed out that our model is different from that in Ref. [22], which deals with another kind of controllable dark hollow beams with rectangular symmetry.

**4. An example**

As an application example, we study the evolution of the intensity distribution of HGSMBs passing through the free space in order to examine the influence of the degree of coherence on its propagation properties. The transfer matrix elements of the free space of distance $z$ are given by $A_1 = A_2 = 1$, $B_1 = B_2 = z$, $C_1 = C_2 = 0$, and $D_1 = D_2 = 1$. In all the following simulations we take the parameters as follows: $\sigma_I = 1$ mm, $\lambda = 632.8$ nm, and the Rayleigh distance $z_R = \pi\sigma_I^2/\lambda$. Figure 1 and 2 show the typical effect of spatial coherence on the evolution of the intensity distributions of the different HGSMBs for the two cases of $n = 1$ and $n = 5$. It is clearly seen that for the beam with large spatial coherence, the hollow Gaussian beam could keep the dark region for a long distance; however, as the decreasing of the spatial coherence, the dark region of the beam shrinks and becomes smaller and smaller and the divergent angle of the beam becomes larger and larger. It also shows that the higher order HGSMB could keep the dark region in much longer propagating distance, compared with the case of the lower order HGSMB. Thus it indicates that for the practical HGSMB, the spatial coherence will greatly affect the application of the dark hollow beam in manipulating or guiding the cold atoms.

## 5. Conclusion

In summary, we have introduced a new partially coherent HGSMB model to describe the practical dark hollow beams. Based on the theory of coherence, a general propagating formula of the HGSMB passing through a paraxial *ABCD* optical system has been analytically obtained. Based on the derived formula, we have given out an application example to illustrate the influence of spatial coherence on the propagation properties of beams. We find that the beam propagating properties of HGSMBs will be greatly affected by their spatial coherence. Our results provide a very convenient way for analyzing the propagation properties of partially coherent dark hollow beams and can be used to analyze the optical trapping or guiding for manipulating atoms and micro-sized particles.

## Acknowledgments

This work has been supported by the National Nature Science Foundation of China (Grant No. 10604047), Scientific Research Foundation from Zhejiang Province (G20630).

**Reference:**

[1] J. Yin, W. Gao, and Y. Zhu, "Generation of dark hollow beams and their applications," in *Progress in Optics*, E. Wolf, ed., North-Holland, vol. **44** (2003) 119.

[2] G. Timp, R. E. Behringer, D. M. Tennant, J. E. Cunningham, M. Prentiss, and K. K. Berggren, Phys. Rev. lett.**69** (1992) 1636

[3] J. P. Yin, Y. F. Zhu, W. B. Wang, Y. Z. Wang, and W. H. Jhe, J. Opt. Soc. Am. B, **15** (1998) 25.

[4] T. Kuga, Y.Torii, N.Shiokawa, and T.Hirano, Phys. Rev. Lett.**78** (1997) 4713.

[5] Y. Cai, X. Lu, and Q. Lin, Opt. Let. **28** (2003) 1084.

[6] R. Grimm, M.Weidemüller, and Y. B. Ovchinnikov, Adv. At. Mol. Opt. Phys. **42** (2000) 95.

[7] M. Kasevich, and S.Chu, Phys.Rev.Lett, **67** (1991) 181.

[8] Y. Song, D. Milam, and W. T. Hill III, Opt. Lett. **24** (1999) 1805.

[9] J. P. Yin, W. J. Gao, H. F. Wang, Q. Long, and Y. Z. Wang, Chinese physics **11** (2002) 1157.

[10] C. Zhao, L. Wang, and X. Lu, Phys. Lett. A **363** (2007) 502.

[11] X. Wang, and M. G. Littman, Opt. lett.**18** (1993) 767.

[12] L. Zhang, X. Lu, X. Chen, and S. He, Chin. Phys. Lett. **21** (2004) 298.

[13] N. R. Heckenberg, R. Mcduff, C.P. Smith, and A.G. White, Opt. Lett. **17** (1992) 221.

[14] A. V. Mamaev, M. Saffman, and A.A. Zozulya, Phys. Rev. Lett. **77** (1996) 4544.

[15] J. Yin, H. Noh, K. Lee, K. Kim, Y. Wang, and W. Jhe, Opt. Commun. **138** (1997) 287.

[16] M. de Angelis, L. Cacciapuoti, G. pierattini, and G. M. Tino, Optics and Laser in Engineering **39** (2003) 283.

[17] D. Ganic, X. Gan, M. Gu, M. Hain, S. Somalingam, S. Stankovic, and T. Tschudi, Opt. Lett. **27** (2002) 1351.

[18] M.A.Clifford, J.Arlt, J. Courtial, and K. Dholakia, Opt. Commun.**156** (1998) 300.

[19] J.Arlt, T. Hitomi, and K. Dholakia, Appl. Phys. B **71** (2000) 549.

[20] J. Arlt, and K. Dholakia, Opt.Commun. **177** (2000) 297.

[21] M. Zhang, and D. Zhao, J. Opt. Soc. Am. A, **22** (2005) 1898.

[22] Y. Cai, and L. Zhang, J. Opt. Soc. Am. B **23** (2006) 1398.

[23] R. Gase, J. Mod. Opt. **38** (1991) 1107.

[24] Q. Lin, and L. Wang, J. Mod. Opt. **50** (2003) 743.

[25] L. Mandel, and E. Wolf, *Optical coherence and quantum optics*, Cambridge University Press 1995.

[26] S. A. Collins, J. Opt. Soc. Am. **60** (1970) 1168.

[27] Q. Lin, and L. Wang, Opt. Commun. **185** (2000) 263.

[28] D. Zwillinger, *CRC Standard Mathematical Tables & Formulae 30$^{th}$ ed.*, CRC press, Beijing, 1998.

[29] A. Erdelyi, W. Magnus, and F. Oberhettinger, *Tables of Integral Transforms*, McGraw-Hill, New York, 1954.

[30] H. M. Srivastava, and J. P. Singhal, Proceedings of the American Mathematical society, **31** (1972) 135.

# FIGURE CAPTIONS

Fig.1: The relative intensity distributions of partially coherent HGSMBs propagating through the free space in the $x-z$ plane with different spatial coherence (a) $\sigma_g/\sigma_I = 10$, (b) $\sigma_g/\sigma_I = 1$, (c) $\sigma_g/\sigma_I = 0.5$ and (d) $\sigma_g/\sigma_I = 0.1$. Here we take the order $n = 1$.

Fig.2: The relative intensity distributions of partially coherent HGSMBs propagating through the free space in the $x-z$ plane with different spatial coherence (a) $\sigma_g/\sigma_I = 10$, (b) $\sigma_g/\sigma_I = 1$, (c) $\sigma_g/\sigma_I = 0.5$ and (d) $\sigma_g/\sigma_I = 0.1$. Here we take the order $n = 5$.

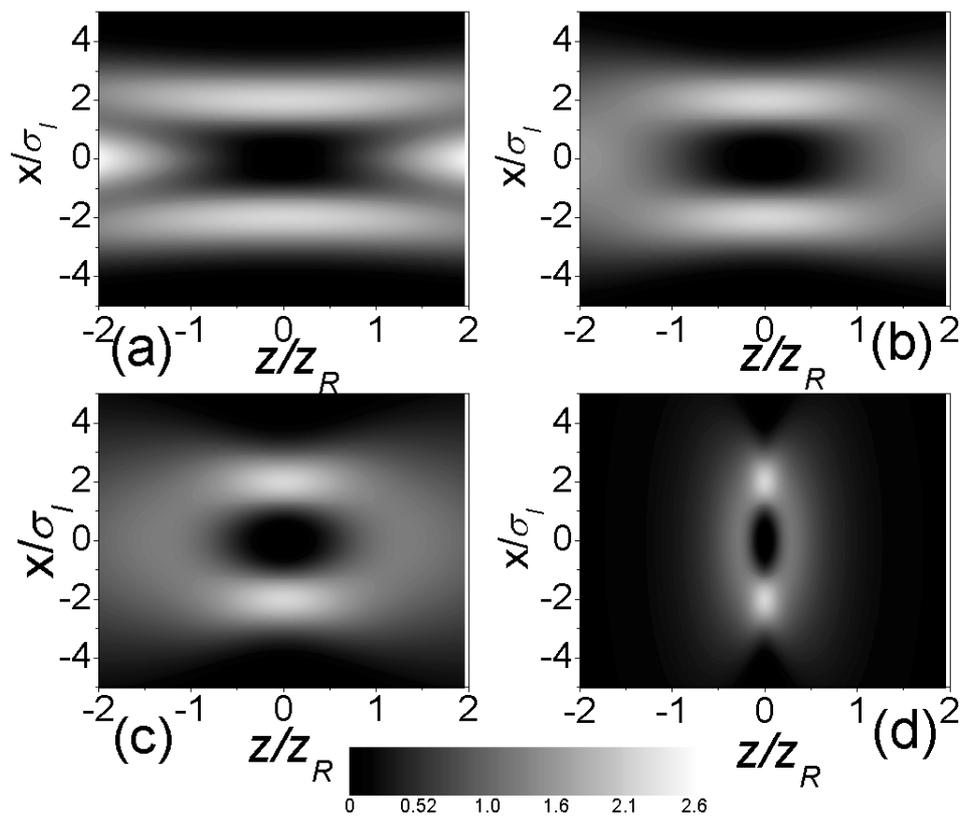

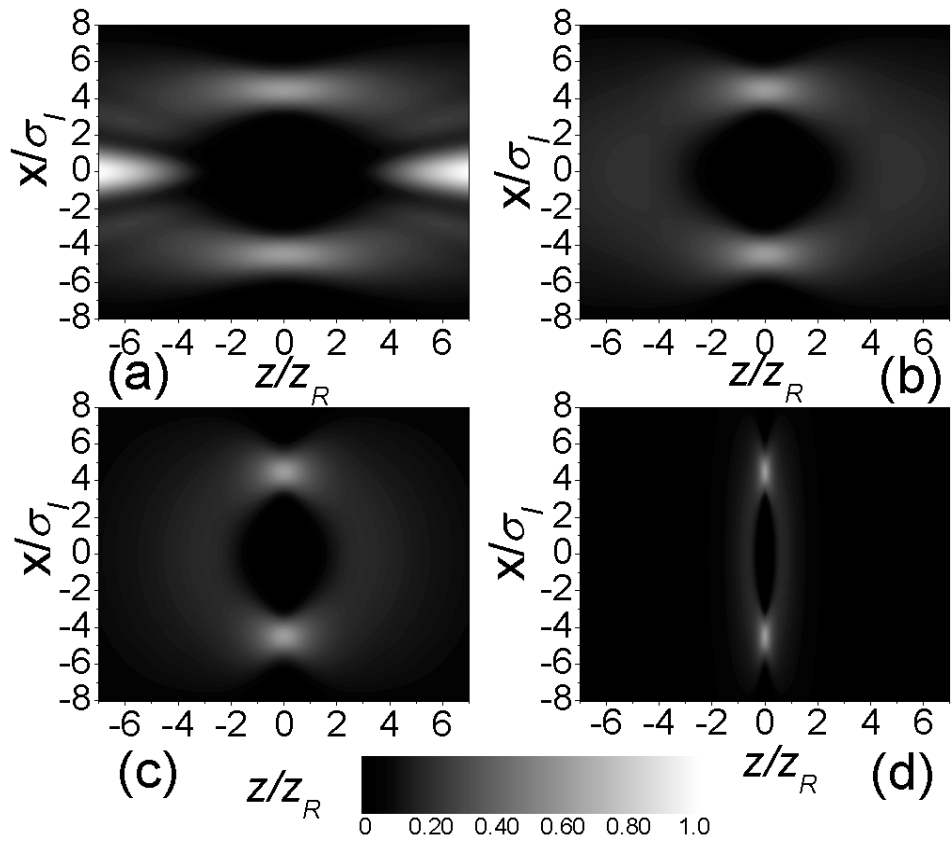